\documentstyle[aps,prl,multicol,epsfig]{revtex}

\begin{document}
\draft
\widetext

\title{
The Unusual Superconducting State of Underdoped Cuprates
}

\author{
Patrick A. Lee and Xiao-Gang Wen
}
\address{Department of Physics, Massachusetts Institute of
Technology,  Cambridge, Massachusetts 02139}

\date{February, 1997}
\maketitle

\widetext
\begin{abstract}
\leftskip 54.8pt
\rightskip 54.8pt

There is increasing experimental evidence that the superconducting energy
gap $\Delta_0$ in the
underdoped cuprates is independent of doping concentration $x$ while the
superfluid density is
linear in $x$.  We show that under these conditions, thermal excitation of
the quasiparticles is
very effective in destroying the superconducting state, so that $T_c$ is
proportional to
$x\Delta_0$ and part of the gap structure remains in the normal state.  We
then estimate $H_{c2}$
and predict it to be proportional to $x^2$.  We also discuss to what extent
the assumptions that
go into the quasiparticle description can be derived in the U(1) and SU(2)
formulations of the
$t$-$J$ model.

\end{abstract}


\begin{multicols}{2}

\narrowtext

While the anomalous properties of cuprate superconductors have been
discussed from the
very beginning, it has become clear in the past several years that it is in the
underdoped region that the cuprates deviate most strongly from conventional
materials,
both in the normal and superconducting states.  NMR,$^{1}$ neutron$^{2}$
and $c$-axis optical
conductivity$^{3}$ indicate the existence of a pseudogap in the normal
state, and
photoemission experiments$^{4,5}$ reveal that the pseudogap is of the same
size and
$\mbox{\boldmath $k$}$ dependence as the superconducting gap.  Furthermore, as
the doping concentrating $x$ is reduced from optimal doping, the
superconducting gap is
constant or may be slightly increasing, while the transition temperature
$T_c$ is
reduced.  Thus a strong deviation from the BCS ratio between energy gap and
$T_c$ is
to be expected.  At the same time, London penetration depth and optical
conductivity$^{6}$
show that the Drude weight in the normal state as well as the superfluid
density in
the superconducting state are proportional to $x$, and a linear relation
between $T_c$
and the superfluid density have been noted.$^{7}$  The small superfluid
density has led a
number of authors to suggest that phase fluctuation may be the determining
factor of
$T_c$ and that the pseudogap may be interpreted as superconducting
fluctuations.$^{8,9}$  A
related interpretation in terms of Bose condensation of pairs has also been
suggested.$^{7,10}$  A second school of thought starts with strongly
correlated models such as
the $t$-$J$ model and interprets the pseudogap as the spin excitation gap
in some RVB
singlet state.  In particular, this scenario is realized in a decomposition
of the
electron into a fermion which carries spin and a boson which represents a
vacancy in
order to enforce the constraint of no double occupation of the $t$-$J$
model.  At the
mean field level, spin and charge are separated, and in the underdoped
region of the
mean field phase diagram the fermions are paired in some intermediate
temperature
range, and become a $d$-wave superconductor only below the Bose condensation
temperature of the bosons.$^{11,12}$  Very recently, a modification of this
mean field theory
was proposed, which incorporates an SU(2) symmetry known to be important at half
filling.$^{13}$  It was argued that the new SU(2) formulation allows a
smoother connection to
half-filling and includes low lying fluctuations ignored in the original U(1)
formulation.  Features in the photoemission data are qualitatively
explained by this
approach, but so far discussions have been limited to the normal state.

In this paper we examine the superconducting state of the underdoped
cuprates.  We
begin with a phenomenological approach, based on the existence of well defined
quasiparticles in the superconducting states.  We show that the combination
of a large
energy gap and small superfluid density leads to unusual features.  A key
result is
that the superconducting state is destroyed by the thermal excitation of
just the low
lying quasiparticles, leaving the large energy gap intact.  We believe this
is a more
effective mechanism of destroying superconductivity than the phase fluctuation
scenario.  With very few assumptions we derive expressions for the temperature
dependence of $\rho_s$ and for $T_c$ and $H_{c2}$.  We then consider whether the
quasiparticle approach can be derived from the microscopic treatments of
the $t$-$J$
model.  We conclude that the U(1) formulation fails to obtain the correct
temperature
dependence of $\rho_s$, even if gauge fluctuations are included in the Gaussian
approximation.  We then indicate how the assumptions of the phenomenology can be
derived from the SU(2) formulation, provided that quantum fluctuations of
low lying
excitations are taken into account.

We limit our discussion to clean superconductors and assume that the elementary
excitations in the superconducting state are well defined quasiparticles with
dispersion
$
E(\mbox{\boldmath $k$}) = \left(
        (\varepsilon_{\mbox{\boldmath $k$}}-\mu)^{2} +
\Delta^{2}_{\mbox{\boldmath $k$}}
        \right)^{1/2}
$
where
$
\Delta_{\mbox{\boldmath $k$}} = \frac{1}{2} \Delta_{0}
(\cos k_x a + \cos k_y a)
$
is a $d$-wave gap with a maximum of $\Delta_{0}$ at $(0, \pi)$.  In a tight
binding
parametrization, $\varepsilon_{\mbox{\boldmath $k$}} = 2t_{f} (\cos k_{x} a
+ \cos k_{y} a)$.  There are 4
nodal points.  In the vicinity of the nodes near $ \left( \frac{\pi}{2},
\frac{\pi}{2}
\right)$, we have the anisotropic Dirac spectrum
$
E(k) = (v_{f}^{2}k_{1}^{2} + v_{2}^{2}k_{2}^{2})^{1/2}
$
where
$
k_1 = \left(
     k_x + k_y - \pi/a
     \right) /
      \sqrt{2}
$,
$k_2 = (k_x - k_y)/ \sqrt{2}$, $v_2 = \frac{1}{\sqrt{2}} \Delta_0 a$, and
$v_f \approx 2
\sqrt{2} t_f a$ for $\mu \approx 0$.  We now make the key assumption that in the
presence of an electromagnetic field the quasiparticle spectrum is shifted
according to

\begin{equation}
E(\mbox{\boldmath $k$},\mbox{\boldmath $A$}) = E(\mbox{\boldmath $k$}) +
\frac{e}{c} \mbox{\boldmath $v$}_{\mbox{\boldmath $k$}} \cdot
\mbox{\boldmath $A$}
\end{equation}

\noindent
where $\mbox{\boldmath $v$}_{\mbox{\boldmath $k$}} = d \varepsilon_{k}/d
\mbox{\boldmath $k$}$ is the normal state velocity.  Equation (1) is correct
to first order in $\mbox{\boldmath $A$}$
in the BCS theory.  It has the consequence that the current carried by the
quasiparticle is $-cdE/d\mbox{\boldmath $A$} = -e\mbox{\boldmath $v$}_{k}$,
i.e. it is the same as the current in the normal state and is different
from the group
velocity $dE_{k}/d\mbox{\boldmath $k$}$.  The difference arises from the fact
that the superconducting quasiparticle is a superposition of particle and hole
states.  For high T$_c$ superconductivity, we do not have the equivalence of BCS
theory, so Eq. (1) must be regarded as an assumption, albeit a very
reasonable one.
This is particularly so near the node, where it is reasonable to believe
that at the
node the superconducting quasiparticle should be the same as the normal
quasiparticle,
so that its current should be given by $\mbox{\boldmath
$v$}_{\mbox{\boldmath $k$}}$.

We next argue that the superfluid tensor defined by $\mbox{\boldmath
$j$}_\mu = \frac{e}{c}
\frac{\rho^{s}_{\mu\nu}}{m} A_\nu$ can be written as $\frac{1}{m}
\rho^{s}_{\mu\nu} =
\frac{1}{m} \rho^{s}(T=0) \delta_{\mu\nu} - \frac{1}{m} \rho^{n}_{\mu\nu}$
where $\rho^{s}(T=0)/m =
x/a^{2}m$ is directly measured by the weight of the Drude peak in the
normal state and by
$\lambda^{-2}_{L}$ where $\lambda_{L}$ is the Landau penetration depth in
the superconducting
state.  By taking $\lambda_{L} = 1600 {\rm\AA}$ for YBCO$_{6.95}$ and $x =
0.2$, we find $m = 2.1
m_e$.  It is convenient to fit this mass to the bottom of a tight binding
bandwidth with hopping integral $t_h$, and
we have $t_h = (2
ma^2)^{-1} = 0.122$ eV which happens to be very close to $J$.  On the other
hand, $\frac{1}{m}\rho_{\mu\nu}^{n}$ is given by the
quasiparticle response to the $\mbox{\boldmath $A$}$ field.  This we can
calculate by writing the free
energy in terms of non-interacting quasiparticles, i.e.,

\begin{equation}
F(\mbox{\boldmath $A$},T) = -kT \sum_{\mbox{\boldmath $k$},\sigma} \ell n
\left(
1+e^{-\beta E(\mbox{\boldmath $k$},\mbox{\boldmath $A$})}
\right)
\end{equation}

\noindent
and differentiating twice with respect to $\mbox{\boldmath $A$}$.  We note
that the neglect of
quasiparticle interaction is justified in the limit of small $T$ and
$\mbox{\boldmath $A$}$
because the density of states of quasiparticles vanishes linearly with
energy, in contrast to the
case of a Fermi liquid.  As explained by Leggett,$^{14}$ the Landau
parameters enter in the form of mean
field theory in Fermi liquid theory.  The vanishing of the free
quasiparticle response functions
implies that the Landau parameters play no role in this limit.  We
therefore obtain

\begin{equation}
\frac{1}{m}\rho_{\mu\nu}^{n} = -2 \sum_{\mbox{\boldmath $k$}}
\frac{dE}{dA_{\mu}}
\frac{dE}{dA_{\mu}} \frac{\partial f}{\partial E} \;\;\;\; .
\end{equation}

\noindent
Strictly speaking, there is an additional term of the form
$
2 \sum_{\mbox{\boldmath $k$}}(\partial^{2}E/\partial A_{\mu}\partial
A_{\nu}) f(E)
$.
This term vanishes in BCS theory due to particle-hole symmetry and we shall
assume that it is also
negligible in the present case.  Using Eq. (1), we replace $dE/dA_{\mu}$ by
the normal state
velocity $\mbox{\boldmath $v$}_{\mu}$ and $\rho_{s}(T)$ can be evaluated in
a straightforward
way

\begin{equation}
\frac{\rho^{s}}{m}(T) = \frac{x}{a^2m} - \alpha T
\end{equation}

\noindent
where
$
\alpha = (2 \ell n (2)/\pi) v_F/v_2 = (8 \ell n(2)/\pi) t_f /\Delta_0
$.
We see that for small $x$, the quasiparticle excitation is an effective way
of destroying the
superconducting state by driving $\rho^s$ to zero.  By extrapolating Eq.
(3) to $\rho^s = 0$, we
can estimate $T_c$ as

\begin{equation}
kT_c \approx 0.88 x \Delta_0 (t_h/t_f) \;\;\;\;\; .
\end{equation}

\noindent
If we assume that $\Delta_0$ is independent of $x$ for underdoped cuprates,
we see that $T_c$ is
proportional to $x$ (or more precisely to $\rho_s (T=0)/m$), thus providing
an explanation of
Uemura's plot.$^{7}$  We shall see that $t_h/t_f \approx 1.8$, so that the
absolute value of $T_c$ given
by Eq. (4) is not unreasonable.  In particular, it is lower than the
estimates based on phase
fluctuation or Bose condensation, which typically gives $T_c \approx xt_h$.
We emphasize that our
mechanism is completely different from these other pictures, in that the
quasiparticle spectrum and
the energy gap $\Delta_0$ comes into play.  Obviously, Eq. (4) implies a
strong deviation from the
BCS ratio between $T_c$ and $\Delta_0$ for small $x$.

Another important implication is that superconductivity is destroyed when
only a small fraction of
the quasiparticles (with energy $\leq x \Delta_0$) are thermally excited.
Thus the gap near
$(0,\pi)$ must remain intact in the normal state, leaving a strip of
thermal excitations which
extend a distance proportional to $x$ from the nodal points.  This is
qualitatively in agreement
with the photoemission experiment.  Of course our phenomenological picture
does not provide a
description of the normal state.  It simply states that the normal state
gap is an inescapable
consequence of a finite $\Delta_0$ and a vanishingly small superfluid
density as $x \rightarrow
0$.

The fact that $d\rho_s/dT$ is independent of $x$ and that both $\rho_s$ and
$T_c$ are proportional
to $x$ means that a scaled plot of $\rho_s(T)/\rho_s(0)$ vs $T/T_c$ should
be independent of $x$
for small $T/T_c$.  In fact, such a scaled plot for YBCO$_{6.95}$ and
YBCO$_{6.60}$ shows a
remarkable universality over the entire temperature range.$^{15}$  We can
use the data to extract the ratio $v_F/v_2$ or
$t_f/\Delta_0$ using Eq. (3).  Using the YBCO$_{6.95}$ data, we obtain a
velocity anisotropy $v_F/v_2 = 6.8$.  A slightly smaller
ratio (by about 15\%) as obtained from the YBCO$_{6.60}$ data.  This
implies that $t_f/\Delta_0 =
1.7$.  If we assume $\Delta_0 = 40$ meV, we find $t_f = 68$ meV, so that
$t_h/t_f = 1.8$ as
mentioned earlier.  Our value of $t_f$ implies a half-filled bandwidth
$\sim 4 t_f \approx 270$ meV
which is consistent with the photoemission data.  This gives $v_f = 1.18
\times 10^7$ cm/sec.

Equation (1) implies that in the presence of a magnetic field, the
quasiparticle spectrum is
shifted so that some of the quasiparticles are  occupied in the ground
state and a finite density
of states is generated at the Fermi energy.  This effect was predicted by
Volovik$^{16}$ and observed in
specific heat measurements.$^{17}$  We shall now use a similar picture to
estimate the size of the vortex
core and estimate H$_{c2}$.  The idea is that as we approach the vortex
core, according to Eq. (2),
the local superfluid density is reduced.  We identify the core size as the
point when $\rho_s$ is
reduced to zero, i.e., when the critical current is reached.  Let us
approach the core in the
$\hat{x}$ direction.  The $\mbox{\boldmath $A$}$ field is in the $\hat{y}$
direction and
$2e\mbox{\boldmath $A$}/c$ should be replaced by the gauge invariant current
$\mbox{\boldmath $\nabla$} \theta - \frac{2e}{c} \mbox{\boldmath $A$}$.
Near the vortex
core the $\mbox{\boldmath $\nabla$} \theta$ term dominate so that we can
approximate
$e\mbox{\boldmath $A$}/c$ by $(2R)^{-1}\hat{y}$.  At $T = 0$ we combine Eq.
(1) and (2) and use
$-df/dE = \delta (E)$ to show that the local normal fluid density at the
point $R\hat{x}$ is given
by

\begin{equation}
\frac{\rho^n}{m} (R\hat{x}) = 4 \left( \frac{v_f}{\sqrt{2}} \right)^2 N \left(
\frac{v_f}{2\sqrt{2}R} \right)
\end{equation}

\noindent
where
$
N(E) = \sum_k \delta \left( E -E(\mbox{\boldmath $k$}) \right)
$
is the quasiparticle density of states for one node.  It is easy to show
that $N(E) = E/(2\pi
v_fv_2)$.  By setting Eq. (5) equal to $\rho^s(T=0)/m$, we estimate the
core size in the $\hat{x}$
direction to be

\begin{equation}
R_1 = \frac{1}{x} (v_f/\pi \Delta_0) \left( t_f/\sqrt{2} t_h \right)
\end{equation}

\noindent
Note that it is greater than the BCS coherence length $v_f/\pi \Delta_0$ by
the factor $x^{-1}$.
On the other hand, using the $T_c$ estimate in Eq. (4), we can write

\begin{equation}
R_1 = v_F/(1.6 \pi kT_c) \;\;\;\; .
\end{equation}

\noindent
which is quite close to the BCS coherence length written in terms of T$_c$
instead of $\Delta_0$.  The two ways of writing the
coherence length are of course equivalent in BCS theory, but very different
for underdoped cuprates.  A main conclusion of this
paper is that Eq. (8) is the proper expression for the coherence length.

 When the core is approached from the (1,1) direction,
similar considerations show that the core size is given by $\sqrt{2} R_1$.
Thus the core takes on an approximately square shape.
This anisotropy in fact extends over long distances and is responsible for
the tendency to form square vortices in
$d$-wave superconductors.$^{18}$  We estimate $H_{c2}$ by assuming that the
square vortex cores are
closed-packed, so that

\begin{equation}
H_{c2} = (hc/2e)/4R_{1}^{2}
\end{equation}

\noindent
Using Eq. (7) and the parameters we extract from the penetration depth data
and setting $\Delta_0 = 40$ meV, $x$ = 0.125 for
YBCO$_{6.6}$, we estimate $R_1 \approx 18.5$ $\rm{\AA}$ and $H_{c2} \approx
146$ T.  Due to the crudeness of the extrapolation
process, we expect both the $T_c$ and $H_{c2}$ expressions to be
overestimates.  Perhaps a more realistic estimate may be made by
using Eq. (8) for $R_1$ and using the experimental $T_c$ and we obtain
instead $R_1 \approx 30$ $\rm{\AA}$ and $H_{c2} \approx 56$
T which is close to the measured value of 40T.$^{19}$
While the absolute value of $H_{c2}$ is quite uncertain, the prediction
that $H_{c2}$ is proportional to $x^2$ (or more
accurately to $\rho_s^2(T=0))$, as long as $\Delta_0$ is constant for
underdoped cuprates, should be amenable to experimental test.
The ideal systems to test this correlation are underdoped YBCO or Hg
cuprates, which fall on the Uemura plot so that $\rho_s(T=0)$
is proportional to $T_c$ and can be accurately determined.  In principle
LACO is a good testing ground because $x$ can be varied.
Unfortunately, there are serious disorder effects for $x \leq 0.1$ and the
nominally pure compound $x = 0.15$ is off the Uemura
plot, for reasons which are not presently understood.  Equation (1) breaks
down in the presence of disorder, and the low lying
excitations may even be localized,$^{20}$ so that the results in this paper
are restricted to the clean limit.

Next we comment on whether existing microscopic models can reproduce the
assumptions of the
quasiparticle description.  In the U(1) formulation of the $t$-$J$ model
the normal state in the
underdoped limit is described by $d$-wave pairing of fermions, so that
there exists an energy gap
$\Delta_0$ which remains finite as $x \rightarrow 0$ in the normal
state.$^{11,12}$  Superconductivity is
driven by condensation of bosons and well defined quasiparticles are
formed.  The superconducting
$T_c$ occurs as an energy scale of $4\pi x t_h$ at the mean field level,
and may be suppressed by
gauge fluctuations.$^{21}$  In this theory the superfluid density is given
by the Ioffe-Larkin rule,$^{22}$
$
\rho^s(T)^{-1} = \rho^{s-1}_f + \rho^{s-1}_b
$.
Since the energy gap appears in the fermion spectrum, we expect
$
\rho^s_f(T) = (1-x) - T/\Delta_0
$
while $\rho^s_b \approx x$ with a higher power in T correction.  Then the
U(1) theory predicts
$
\rho^s(T) = x - x^2 T/\Delta_0
$.
While the $T=0$ value is correctly given to be $x$,
the temperature dependence is in strong disagreement with Eq. (3) and with
experiment in that $\alpha$ is suppressed by
$x^2$.  The origin of this difficulty is that the fermion does
not couple directly to $\mbox{\boldmath $A$}$, but to the U(1) gauge field
$\mbox{\boldmath $a$}$ while the bosons couple to $\mbox{\boldmath $A$} +
\mbox{\boldmath $a$}$.  The external $\mbox{\boldmath $A$}$ produces a finite
$\mbox{\boldmath $a$}$ but its magnitude is reduced by $x$.  In the
quasiparticle language,
the shift of the spectrum in the presence of $\mbox{\boldmath $A$}$ is
smaller than that given
in Eq. (1) by $x$.  It is difficult to escape from this conclusion in the
U(1) theory, because
gauge fluctuations are included at the Gaussian level in the Ioffe-Larkin
rule, which should be a
good approximation in the superconducting state in that the gauge mode is
strongly gapped by the
Anderson-Higgs mechanism.

It was shown recently$^{13}$ that the U(1) formulation does not connect
smoothly to the half-filled limit
which is known to exhibit an SU(2) symmetry.  For small $x$, there are
indeed low lying gauge
fluctuations with energy scale of order $x\Delta_0$ which are ignored in
the U(1) formulation.  A
new SU(2) formulation was introduced, which allows these low energy
fluctuations to be described in a natural way.  
The low energy effective theory contains a boson part and a fermion part: 
$L_{eff}=L_b+L_f$.  (For details see Ref. 13.)
The boson part is given by
\begin{eqnarray}
L_{b} &=&
i b^\dagger (\partial_t -ieA_0-ia_0 \tau^3) b 
- \frac{1}{2m} |(\partial_i -i\frac{e}{c} A_i - ia_i \tau^3) b|^2 \nonumber \\
&& - \frac{D_1}{2m} (b^\dagger b)^2 -\mu b^\dagger b
- D_2 \frac{J }{2} ( |b_1|^2- |b_2 |^2)^2
\end{eqnarray}
where $b=\pmatrix{b_1\cr b_2\cr}$, and
$D_{1,2}$ are order one coefficients. The fermions are in a staggered
flux phase and couple only to the $a_\mu$ gauge field.
When $|b_1|=|b_2|\neq 0$, the system is in a superconducting state (which corresponds
to the d-wave paired state in the U(1) formulation).
When $b_1\neq 0$ and $b_2=0$, the system is in a metallic state (which
corresponds to the staggered flux phase in the U(1) formulation).
Since $D_2>  0$ the ground state is the superconducting state.
The normal state at finite temperatures contains no
boson condensation and is a state which fluctuates
between $d$-wave
pairing and the staggered flux phase of fermions.
The fermion spectrum acquires a gap $\Delta_0$
which is finite for small $x$.  

\begin{figure}
\epsfysize=1.1truein
\centerline{ \epsffile{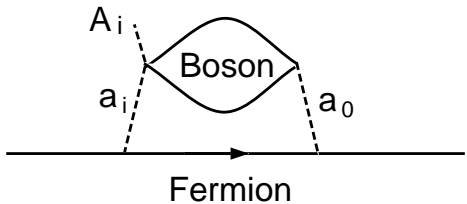} }
\caption{
Coupling between $A_i$ and fermion quasiparticles.
}
\end{figure}

We note that, in the superconducting state ($|b_1|=|b_2|$), $a_\mu$ and 
$A_\mu$ decouple under the meanfield approximation. Therefore the
low lying fermion quasiparticles do not couple to the external electromagnetic
gauge field $A_\mu$, and cannot reduce the superfluid density within meanfield 
theory. However
unlike the U(1) case, quantum fluctuations
of the gauge fields and the quantum fluctuations between the two bosons
(both are omitted in the U(1) formulation) are
important even at $T = 0$.  Those quantum fluctuations
induce a coupling between the fermion quasiparticles
and the gauge potential $\mbox{\boldmath $A$}$. One such contributions
is illustrated in Fig. 1.
We find that the shift
of the quasiparticle
spectrum is of the form given by Eq. (1), except that the $\mbox{\boldmath
$v$}_k \cdot
\mbox{\boldmath $A$}$ is multiplied by a numerical constant of order unity.
Thus the SU(2)
theory incorporates the main ingredients underlying the present paper,
i.e., a finite gap
$\Delta_0$, a superfluid density proportional to $x$, and a quasiparticle
spectrum given by Eq.
(1).  Details of this microscopic theory will be given elsewhere.

We believe that our prediction that $H_{c2} \sim x^2$ is significant for
two reasons.  First, it is
in contrast with models based on Bose condensation which should predict
$H_{c2} \sim x$ since the
coherence length in that case is the interparticle spacing $x^{-1/2}$.
Secondly, $H_{c2} \sim x^2$
is a weak field in the sense that when compared with the hole density $x$,
the number of Landau
levels occupied is $x^{-1} \gg 1$ so that we are outside of the quantum
Hall limit.  Thus
we expect the state for $H > H_{c2}$ to be a metallic state and the key
question is what kind of
metallic state it is.  It is clear from the present discussion that for $x
\ll 1$, the energy gap
at $(0,\pi)$ survives inside the vortex core and therefore for $H >
H_{c2}$.  The magnetic field
drives a region of gapless excitations in the Brillouin zone which extends
a distance $x$ from the
nodal positions, qualitatively similar to the normal state above $T_c$.  It
seems to us that two possibilities remain.  First the
gapless excitations are well defined quasiparticles in the Landau sense.
In this case, Luttinger theorem requires that a
breaking of translation symmetry must occur to produce the energy gap and
the metallic state may be
understood as some form of staggered flux phase.  Alternatively, the
gapless excitations are not
Landau quasiparticles, but acquire residual width due to quantum
fluctuations, making this state a
genuine non-Fermi liquid state.  This latter scenario is an exciting
possibility which deserves further investigation.

We thank G. Boebinger, J. Cooper, J. Loram and T. Timusk for important
discussions.  PAL acknowledges support by NSF-MRSEC Grant
No. DMR--94--00334.  XGW is supported by NSF Grant No. DMR--94--11574.

\end{multicols}

\end{document}